\documentclass[final]{IEEEtran}
\usepackage{amsthm,amssymb,graphicx,multirow,amsmath,color,amsfonts}
\usepackage[update,prepend]{epstopdf}
\usepackage[noadjust]{cite}
\usepackage[latin1]{inputenc}
\usepackage{tikz}
\usetikzlibrary{arrows,calc}		
\usepackage{bbm} 
\usepackage{pdfpages}
\usepackage{tabulary}
\usepackage{multirow}
\usepackage{comment}
\usepackage{graphicx}
\usepackage[font=small,labelfont=bf]{caption}
\usepackage{subcaption}
\usepackage[makeroom]{cancel}
\usepackage{microtype}
\usepackage{xcolor,etoolbox}

\usepackage{setspace}	

\def\nb0{{\mathbf{0}}}
\def\nb1{{\mathbf{1}}}

\newtheorem{lemma}{Lemma}

\newtheorem{theorem}{Theorem}

\newtheorem{remark}{Remark}

\allowdisplaybreaks 

\newcommand\cc[1]{\color{black}#1}

\makeatletter 
\pretocmd\@bibitem{\color{black}\csname keycolor#1\endcsname}{}{\fail}
\newcommand\citecolor[1]{\@namedef{keycolor#1}{\color{blue}}}
\makeatother

\begin{document}
\graphicspath{{./Figures/}}
\title{Manchester Encoding for Non-coherent Detection of Ambient Backscatter in Time-Selective Fading
}
\author{ 
J. Kartheek Devineni, \emph{Graduate Student Member, IEEE}, and Harpreet S. Dhillon, \emph{Senior Member, IEEE} 
\thanks{Copyright (c) 2021 IEEE. Personal use of this material is permitted. However, permission to use this material for any other purposes must be obtained from the IEEE by sending a request to pubs-permissions@ieee.org.} 
\thanks{J. K. Devineni and H. S. Dhillon  are with Wireless@VT, Department of ECE, Virginia Tech, Blacksburg, VA (email: kartheekdj@vt.edu and hdhillon@vt.edu). This work was supported by U.S. NSF Grants CNS-1814477 and CPS-1739642.
} 
}

\maketitle
\vspace{-1.5cm}
\begin{abstract}

Simplifying the detection procedure and improving the bit error rate (BER) performance of a non-coherent receiver in ambient backscatter is vital for enhancing its ability to function without the channel state information (CSI). In this work, we analyze the BER performance of Manchester encoding which is implemented at the transmitter for data transmission, and demonstrate that the optimal decision rule is independent of the system parameters. {\cc Further, through extensive numerical results, it is shown that the ambient backscatter system can achieve a signal-to-noise ratio (SNR) gain with Manchester encoding compared to the commonly used uncoded direct on-off keying (OOK) modulation, when used in conjunction with a multi-antenna receiver employing the direct-link cancellation.}

\end{abstract}


\section{Introduction} \label{sec:intro}

The potential applications of ambient backscatter for the next generation wireless networks necessitates the technology to support information transfer not only among heterogeneous user equipment (UE), but also under diverse channel scenarios. For example, the emerging paradigm of smart cities will require ambient backscatter to perform well in both the slow and fast varying channels. While there have been a lot of studies on slow fading channels in the literature \cite{devineni2019ambient}, the investigation into time-selective fading channels has not received enough attention. Due to the difficulty of acquiring channel state information (CSI) for such fast varying channels, it might be detrimental to build a system based on coherent communication, and could be more beneficial to choose non-coherent transmission as the preferred mode of communication in such channels. It is, therefore, very important to improve the performance of non-coherent detection for the ambient backscatter to accelerate its widespread adoption and implementation for applications that experience  fast varying channels, such as the vehicular communications systems. Towards this goal, we investigate the advantages of employing  Manchester encoding for the non-coherent transmission of ambient backscatter symbols under a time-selective fading setup. We show that this encoding scheme can result in reduction of detection complexity at the receiver, while improving the BER performance compared to the popular on-off keying (OOK) modulation. 

\emph{Related Work:} With respect to the assumptions about the channel model, the current literature on non-coherent ambient backscatter can be broadly divided into two categories. The first category belongs to slow fading channels, for which the non-coherent receiver designs based on maximum-likelihood (ML) detection \cite{gao17}, semi-coherent detection \cite{gao16}, and orthogonal frequency division multiplexing (OFDM) \cite{el2019noncoherent, elmossallamy2019noncoherent} are proposed. Also, the blind channel estimation techniques that do not require transmission of separate pilot signals are studied in \cite{zhao2018blind, guo2018noncoherent} for the ambient backscatter setup. Manchester encoding was first explored in \cite{tao2018symbol} for a slow fading ambient backscatter setup. {\cc In \cite{zhao2019channel}, the angle of arrival estimation (AoA) using a reader with a massive number of antennas is explored for the ambient backscatter setup.} The second category relates to the time-selective fading channels, which is of more interest to us but has not received much attention. The most relevant prior art in this direction is our own work \cite{devineni2019non}, which focuses on the non-coherent multi-antenna receiver design for direct OOK modulation. However, the bit error rate (BER) analysis of the ambient backscatter systems under Manchester encoding and time-selective fading is an open problem, which is solved in this paper. A particular technical novelty of the paper is in carefully handling the correlation between the test statistics corresponding to the two codewords of the encoded symbol, which is crucial for the exact BER analysis. 

\emph{Contributions:} {\cc In this work, we introduce Manchester encoding to the time-selective fading setup of an ambient backscatter system, and analyze the performance of the scheme. We also determine the advantages of Manchester encoding over the direct OOK modulation by comparing the complexity of the two detection mechanisms and their BER performance.} A low-complexity receiver architecture based on the direct averaging of the received signal samples is considered for the setup. { \cc This architecture diverges from the conventional one based on the averaging of energy of the received signal samples, which is commonly used in the ambient backscatter literature \cite{devineni2019ambient}.} The main contributions of our current work can be summarized as follows: 1) evaluation of the conditional joint distributions and the average BER of Manchester encoding for both the single antenna (SA) and multi-antenna (MA) receivers,  and 2) novel analysis that demonstrates the advantages of Manchester encoding over the popular direct OOK modulation. {\cc To be exact, we analytically show that the optimal detection rule of the ambient backscatter with Manchester encoding is independent of the system and channel parameters, which greatly simplifies the receiver implementation. In addition, the encoding scheme also results in an SNR gain over the direct OOK modulation, when used in conjunction with an MA receiver implementing the direct-link (DL) cancellation. The exact gain in SNR is dependent on the joint distribution of the AoAs of the DL and backscatter link (BL). For the uniform spread and narrow spread joint distributions of the AoAs considered in our work, the SNR gain comes out to be around $4$ dB and $3$ dB, respectively.}

\section{System Model} \label{sec:SysMod}

\begin{figure}
    \centering
        \includegraphics [width=0.9\linewidth]{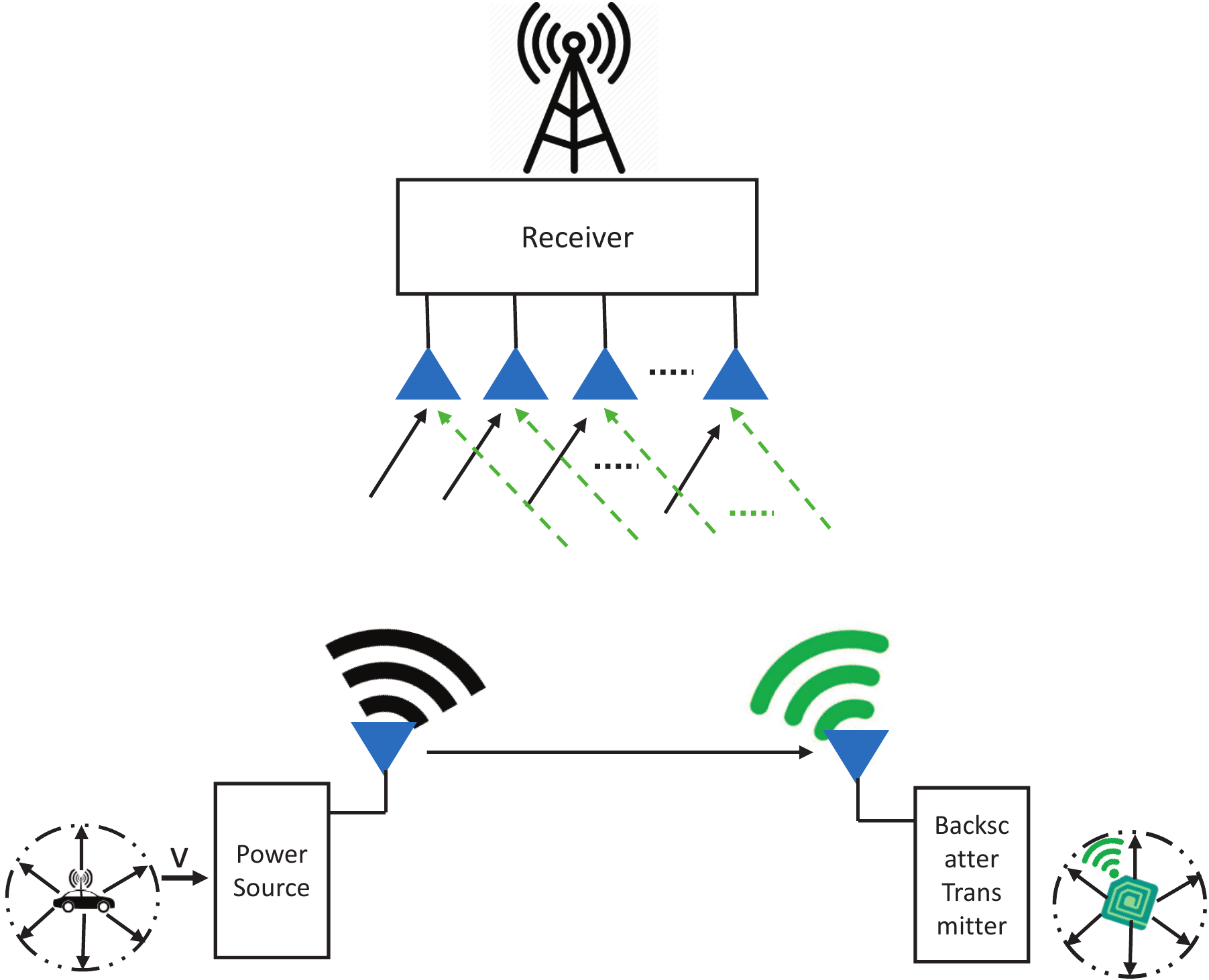}
	\caption{\cc System model for the ambient backscatter setup.}\label{fig:systemfig}
\end{figure}

The setup for the ambient backscatter system mainly consists of three devices, namely the ambient power source (PS), the backscatter transmitter (BTx), and the receiver (Rx). The ambient PS and BTx are surrounded by local scatterers resulting in independent sub-paths with uniformly distributed angle of departure (AoD), while the Rx only has dominant scatterers that are far away, resulting a narrow spread for the AoA. Hence, the signal at Rx can be modeled as spatially correlated with two receive links, namely DL and BL that arrive at AoAs $\theta_1$ and $\theta_2$, respectively, after propagating through a flat Rayleigh fading channel. {\cc The described system setup of the ambient backscatter is illustrated in Fig. \ref{fig:systemfig}.} In addition, the ambient PS and BTx could be in motion independently of each other, resulting in a time-varying channel. {\cc Under the local scattering assumptions, the auto-correlation function (ACF) for the fading process of the DL and BL links is given by $J_0(2\pi f_d t_d)$, where $J_0()$ is the zero order Bessel function of the first kind, $f_d$ is the maximum Doppler spread (DS) of the link and $t_d$ is the delay \cite{stuber2017principles}.} Similarly, the ACF for the PS-BTx link is given by $J_0(2\pi f_d t_d) J_0(2\pi a f_d t_d)$, where $a$ is the ratio of the DS present at these two devices of the link \cite{stuber2017principles}. For tractability, the temporal fading of each link is modeled as a first-order AR process given by $ h[n] \!= \! \rho  h[n-\!1] + \sqrt{1\!-\!\rho^2} g[n]$, where $h[n]$ and $h[n\!-\!1]$ are the gains of the current and previous time instants, respectively, $g[n]$ is the complex Gaussian process of variance $\sigma_h^2$, and $\rho \in [0, 1)$ is the correlation factor \cite{liu2002space}.  Depending on the link, the value of $\rho$ is given by either $J_0(2\pi f_d T_s)$ or $J_0(2\pi f_d T_s ) J_0(2\pi a f_d T_s)$, where $T_s$ is the symbol duration. The MA receiver of the current setup utilizes the slow varying rate of the large scale parameter AoA, in comparison to varying rate of the overall channel gain of the fading channel, to track the AoA of the DL and cancel its interference \cite{devineni2019non}.

By intentionally keeping the data rate of backscatter lower compared to that of the ambient data, the signal at the SA receiver can be expressed as:
\begin{align}
	y[n] &= h_r[n] x[n]+\alpha b \, h_{b}[n]\,  h_t[n] x[n] + w[n], \label{eq:actmod}
\end{align}
where $x[n]$ is the ambient data sequence, $w[n]$ is the additive Gaussian noise, $h_r[n], h_b[n]$ and $h_t[n]$ are i.i.d. zero mean complex Gaussian channel coefficients with variance $\sigma_h^2$ and are unknown at Rx, $b$ is the backscatter data bit, and $\alpha$ is related to the parameter $\Gamma_1$ (the reflection coefficient of the tag when bit `1' is transmitted) of the BTx node. The channel coefficients $h_r[n], h_b[n]$ and $h_t[n]$ are modeled using a first-order AR process, each having a separate correlation factor $\rho_r, \rho_b$, and $\rho_t$, respectively. Similarly, the resultant signal at the MA receiver after the DL cancellation is given by:
\begin{align}
	\tilde{\mathbf{y}}[n] & = \tilde{\mathbf{a}} \, \alpha b \, h_{b}[n] h_t[n]  x[n] + \tilde{\mathbf{w}}[n],  \label{eq:effsignal}
\end{align}
where the resultant vectors $\tilde{\mathbf{a}}$ and $\tilde{\mathbf{w}}[n]$ are given by:
\begin{gather}
\tilde{\mathbf{a}} 
= \begin{bmatrix}
     2 \sin (\frac{\phi_2-\phi_1}{2}) e^{j(\frac{\phi_2-\phi_1}{2})}  \\
    \vdots\\
    2 \sin (M_r-1)(\frac{\phi_2-\phi_1}{2}) e^{j(M_r-1)(\frac{\phi_2-\phi_1}{2})} 
  \end{bmatrix}\!\!,\nonumber\\ 
   \tilde{\mathbf{w}}[n] = \begin{bmatrix}
    e^{- j\phi_1} w_1[n] - w_0[n] \\
    \vdots\\
    e^{- j(M_r-1)\phi_1} w_{M_r-1}[n] - w_0[n] 
  \end{bmatrix}\!\!.
\end{gather}
The phase offset $\phi_i$ of each link is given by $ \frac{2\pi}{\lambda} d \cos \theta_i$. Additional details on the setup, channel and signal model can be found in \cite{devineni2019non}. However, unlike the direct OOK modulation used in \cite{devineni2019non}, the transmitter in this case sends out codewords $[0\, 1]$ and $[1\, 0]$, known as Manchester coding, using the OOK modulation for message ($b$) bits $0$ and $1$, respectively. For completeness, note that a preliminary study of Manchester encoding appears in our conference paper \cite{devineni19noncoherent}, which is limited to a dual-antenna receiver and assumed independent fading across the ambient symbols. For a fair comparison with the direct OOK modulation, we assume that each codeword of Manchester encoding is sent within a single symbol duration of the backscatter data instead of the time duration of two symbols. The test statistics (TSs) $Z_0$ and $Z_1$ are evaluated for the two symbols of the codeword by taking half the samples each from the sample size $N$, and are given by $Z_0 = \frac{2}{N} \sum_{n=1}^{N/2} y[n]$ and $Z_1 = \frac{2}{N} \sum_{n=N/2+1}^{N} y[n]$. Although the setup of the paper is inspired by \cite{devineni2019non}, the new analysis for Manchester encoding under time-selective fading channel is fundamentally different and non-trivial due to correlation between $Z_0$ and $Z_1$.

\section{Detection at the Single Antenna Receiver} \label{sec:DetectnBERSA}

In this section, we evaluate the performance of Manchester encoding in a SA receiver by deriving the conditional probability density functions (PDFs) and BER of the receiver. {\cc The BER probability of the detector is represented using one of the commonly used notations $P(e)$, where $e$ is the bit error event.}

\subsection{Conditional Distributions of the Signal}\label{sec:ConDist}

The null and alternate hypotheses $\mathcal{H}_0$ and $\mathcal{H}_1$ of the encoding scheme correspond to the backscatter bit $b \equiv 0$ and $b \equiv 1$, respectively. Since the transmitter sends out codewords, we have to derive the joint conditional distributions of the TSs $Z_0$ and $Z_1$ evaluated for each symbol of the codeword.

\begin{lemma} \label{lem:CondPDF}
The joint PDFs of $Z_0$ and $Z_1$ conditioned on $\mathcal{H}_0$ and $\mathcal{H}_1$ for Manchester encoding in SA receiver are given by: 
\begin{align}	
	&\mathcal{H}_0 : f_{Z_0, Z_1 }\left( z_0, z_1 \right) \nonumber\\
	&\!=\!  \frac{\exp\left\{\!-\!\left( \frac{\left|z_0\right|^2 {\rm Var}^{\rm SA}_1  + \left|z_1\right|^2 {\rm Var}^{\rm SA}_0 - (z_0 z_1^*+ z_0^* z_1) {\rm Cov}^{\rm SA}}{{\rm Var}^{\rm SA}_0 {\rm Var}^{\rm SA}_1 - \left({\rm Cov}^{\rm SA}\right)^2}\right)\right\}}{\pi^2 \left({\rm Var}^{\rm SA}_0 {\rm Var}^{\rm SA}_1 - \left({\rm Cov}^{\rm SA}\right)^2\right)},  \\
	& \mathcal{H}_1 : f_{Z_0, Z_1 }\left( z_0, z_1 \right) \nonumber\\
	&\!=\!  \frac{\exp\left\{\!-\!\left( \frac{\left|z_0\right|^2 {\rm Var}^{\rm SA}_0  + \left|z_1\right|^2 {\rm Var}^{\rm SA}_1 - (z_0 z_1^*+ z_0^* z_1) {\rm Cov}^{\rm SA}}{{\rm Var}^{\rm SA}_0 {\rm Var}^{\rm SA}_1 - \left({\rm Cov}^{\rm SA}\right)^2}\right)\right\}}{\pi^2 \left({\rm Var}^{\rm SA}_0 {\rm Var}^{\rm SA}_1 - \left({\rm Cov}^{\rm SA}\right)^2\right)}  \label{eq:condpdfM2},
\end{align}
where ${\rm Var}^{\rm SA}_0 = \frac{2(\sigma_h^2 \mathbb{E} \left[ |X|^2 \right] + \sigma_h^2 \frac{2\rho_r}{1-\rho_r} ( 1 - \frac{2(1-\rho_r^{N/2})}{N(1-\rho_r)} ) |\mathbb{E} \left[ X \right]|^2 + \sigma_n^2)}{N}$, 
\begin{gather*}
{\rm Var}^{\rm SA}_1\!\!=\!\! \frac{2}{N} (\sigma_h^2 (1 \!+ \!|\alpha|^2  \sigma_h^2) \mathbb{E} \!\left[ |X|^2 \right] \!+\! \sigma_h^2 \Big[\!\frac{2\rho_r}{1\!\!-\!\!\rho_r} ( 1 \!\!-\!\! \frac{2(1\!\!-\!\!\rho_r^{N/2})}{N(1\!\!-\!\!\rho_r)} )\\
	 +  |\alpha|^2  \sigma_h^2 \frac{2\rho_t \rho_b}{1-\rho_t \rho_b} ( 1 \!-\! \frac{2(1\!-\!\rho_t^{N/2} \rho_b^{N/2})}{N(1\!-\!\rho_t \rho_b)} ) \Big] |\mathbb{E} \left[ X \right]|^2  + \sigma_n^2),
\end{gather*}
	  and ${\rm Cov}^{\rm SA} = \frac{4\rho_r \left( 1-\rho_r^{N/2}\right)^2}{N^2\left( 1-\rho_r\right)^2}  |\mathbb{E} \left[ X \right]\!|^2$.
\end{lemma}

\begin{IEEEproof}
	See Appendix~\ref{app:CondPDF}.
\end{IEEEproof}

\subsection{Bit Error Rate} \label{sec:DetBERSO}

The conditional PDFs of  $Z_0$ and $Z_1$ under the two hypotheses are compared to derive the optimal threshold, which is used to evaluate the BER of the SA receiver. 

\begin{theorem} \label{thm:BERExpSO}
The average BER of Manchester encoding in the SA receiver is given by:
\begin{align}
P_{\rm SA}(e) &= \int_{0}^{\infty} \int_{v}^{\infty} \frac{\exp\left\{-\left( \frac{u}{( 1 - \rho^2){\rm Var}_0^{\rm SA}}  + \frac{v}{( 1 - \rho^2){\rm Var}_1^{\rm SA}} \right)\right\}}{\pi ( 1 - \rho^2){\rm Var}_0^{\rm SA}{\rm Var}_1^{\rm SA}} \nonumber\\
& \quad \quad \quad  \times I_0(\frac{\rho \sqrt{u v}}{( 1 - \rho^2)\sqrt{{\rm Var}_0^{\rm SA}{\rm Var}_1^{\rm SA}}}) \,\mathrm{d}u \,\mathrm{d}v, \label{eq:bersa}
\end{align}
where $I_0$ is zeroth order modified Bessel function of the first kind. The expression in~\eqref{eq:bersa} can be well approximated as \begin{align}
P_{\rm SA}(e) &= \left(\!1\!+\! \frac{{\rm Var}^{\rm SA}_1}{{\rm Var}^{\rm SA}_0}\!\right)^{\!\!-1}
\end{align}
 for large values of the sample size $N$, since the two variances ${\rm Var}^{\rm SA}_0$ and ${\rm Var}^{\rm SA}_1$ both decay at the rate of $\Theta(N^{-1})$ while the covariance ${\rm Cov}^{\rm SA}$ decays at the rate of $\Theta(N^{-2})$.
\end{theorem}

\begin{IEEEproof}
	See Appendix~\ref{app:BERExpSO}.
\end{IEEEproof} 

\subsubsection*{\textbf{Asymptotic analysis}}
The ratio of the variances of the null and alternate hypotheses of the SA receiver is
\begin{align*}
K 
&= 1 + \frac{ |\alpha|^2 \sigma_h^4 \left\{ 1 +  \frac{2\rho_t \rho_b}{1-\rho_t \rho_b} ( 1 \!-\! \frac{2(1\!-\!\rho_t^{N/2} \rho_b^{N/2})}{N(1\!-\!\rho_t \rho_b)} ) \frac{ |\mathbb{E} \left[ X \right]|^2}{\mathbb{E} \left[ |X|^2 \right]}  \right\}}{\sigma_h^2 \left\{1 +  \frac{2\rho_r}{1\!\!-\!\!\rho_r} ( 1 \!\!-\!\! \frac{2(1-\rho_r^{N/2)}}{N(1-\rho_r)} ) \frac{ |\mathbb{E} \left[ X \right]|^2}{\mathbb{E} \left[ |X|^2 \right]}  \right\} + {\rm SNR}^{-1}} .
\end{align*}
The asymptotic BER of Manchester encoding in the SA receiver is given by:
\begin{align*}
&P_{\rm SA}^{\rm asym}(e) = \lim_{{\rm SNR} \to \infty} (1+K)^{-1}\nonumber\\
&= \left(2 \!+\! \frac{|\alpha|^2 \sigma_h^4  \!+\! |\alpha|^2 \sigma_h^4 \frac{2\rho_t \rho_b}{1-\rho_t \rho_b} ( 1 \!-\! \frac{2(1\!-\!\rho_t^{N/2} \rho_b^{N/2})}{N(1\!-\!\rho_t \rho_b)} ) \frac{ |\mathbb{E} \left[ X \right]|^2}{\mathbb{E} \left[ |X|^2 \right]}}{1  + \frac{2\rho_r}{1\!-\!\rho_r} ( 1 \!\!-\!\! \frac{2(1-\rho_r^{N/2)}}{N(1-\rho_r)} ) \frac{ |\mathbb{E} \left[ X \right]|^2}{\mathbb{E} \left[ |X|^2 \right]}} \right)^{\!\!\!\!-1} \!\!\!\!. 
\end{align*}

\begin{remark}\label{rem:DetTh}
It is important to highlight the advantages of  Manchester encoding over the direct OOK modulation. The decision rule, given in~\eqref{eq: optdecrule}, for Manchester encoding is just a function of the relative magnitudes of the two RVs $Z_0$ and $Z_1$. On the other hand, the decision rule of the direct OOK modulation is based on the comparison of the magnitude of a single TS $Z$ (evaluated over all the $N$ samples) with a threshold, which can be expressed as follows:
\begin{align}
|z|^2   \gtrless_{0}^{1}  \ln ( \frac{s_1}{s_0} )\frac{s_1 s_0}{s_1-s_0},
\end{align}
where $s_0 = \frac{(\sigma_h^2 \mathbb{E} \left[ |X|^2 \right] + \sigma_h^2 \frac{2\rho_r}{1-\rho_r} ( 1 - \frac{(1-\rho_r^{N})}{N(1-\rho_r)} ) |\mathbb{E} \left[ X \right]|^2 + \sigma_n^2)}{N}$ and
\begin{gather*}
s_1\!\!=\!\! \frac{1}{N} (\sigma_h^2 (1 \!+ \!|\alpha|^2  \sigma_h^2) \mathbb{E} \!\left[ |X|^2 \right] \!+\! \sigma_h^2 \Big[\!\frac{2\rho_r}{1\!\!-\!\!\rho_r} ( 1 \!\!-\!\! \frac{(1\!\!-\!\!\rho_r^{N})}{N(1\!\!-\!\!\rho_r)} )\\
	 +  |\alpha|^2  \sigma_h^2 \frac{2\rho_t \rho_b}{1-\rho_t \rho_b} ( 1 \!-\! \frac{(1\!-\!\rho_t^{N} \rho_b^{N})}{N(1\!-\!\rho_t \rho_b)} ) \Big] |\mathbb{E} \left[ X \right]\!|^2  + \sigma_n^2).
\end{gather*}
The optimal decision rule for the direct OOK modulation is a function of the system parameters such as the SNR of ambient signal, the fading variance $\sigma^2_h$, the sample size $N$, and the correlation factors $\rho_r, \rho_b$ and $\rho_t$ \cite{devineni2019non}. 
Hence, this scheme  considerably reduces the receiver complexity, and will most likely be preferred for the cases where optimizing the energy consumption of the device is a priority. 
\end{remark}
Even though the optimal decision rule is simplified with Manchester encoding, the asymptotic BER still suffers from an error floor. Therefore, the performance of the encoding scheme in the MA receiver needs to be evaluated to demonstrate its full potential. {\cc In the next section, we discuss the antenna gain, detection procedure and the BER performance of the MA receiver when Manchester encoding is employed at the backscatter transmitter. }

\section{Detection at the Multi-Antenna Receiver} \label{sec:DetectnBERMA}

The effective signal obtained after the DL cancellation, and proper weighting of the resultant signal vector in the MA receiver is given by \cite{devineni2019non}:
\begin{align}
 y_{\rm eff}[n] 
 & = \mathbf{r}^* \tilde{\mathbf{a}} \alpha b \, h_{b}[n] h_t[n]  x[n] + \mathbf{r}^* \tilde{\mathbf{w}}[n]  \label{eq:effsignalMr},
\end{align}
where $\mathbf{r} =  \frac{ \mathbf{\hat{K}_{\tilde{W}}^{-1}} \tilde{\mathbf{a}}}{|\mathbf{\hat{K}_{\tilde{W}}^{-\frac{1}{2}}} \tilde{\mathbf{a}}|}$ is the weight vector with the optimal MMSE detection.
 The antenna gain $ G =\tilde{\mathbf{a}}^* \mathbf{\hat{K}_{\tilde{W}}^{-1}} \tilde{\mathbf{a}}$ due to multiple antennas is given by \cite{devineni2019non}:
\begin{align*}
G &= M_r\!-\!\frac{1}{M_r} \!- \!\frac{2}{M_r} \frac{\sin  ((M_r\!-\!1)\frac{\phi_2-\phi_1}{2} ) }{\sin (\frac{\phi_2-\phi_1}{2} )}  \cos ( M_r\frac{\phi_2-\phi_1}{2}) \nonumber\\
& - \!\frac{1}{M_r} \frac{\sin^2  ((M_r\!-\!1)\frac{\phi_2-\phi_1}{2} ) }{\sin^2 (\frac{\phi_2-\phi_1}{2} )},
\end{align*}
{\cc and the expression can be further simplified as follows:
\begin{align}
G &= M_r  - \frac{1}{M_r} \frac{\sin^2 (M_r \frac{\phi_2-\phi_1}{2})  }{\sin^2 (\frac{\phi_2-\phi_1}{2} )} .
\end{align}
}
Although the antenna gain is a function of the two phase offsets (and thereby the AoAs), it is represented as a single variable $G$ without any arguments to simplify the notation. 

{\cc The exact expression of the average BER is dependent on the joint distribution of the two variables $\theta_1$ and $\theta_2$. For further exposition, we consider two kinds of distributions: 1) {\em uniform spread of AoAs:}  the  two AoAs $\theta_1$ and $\theta_2$ are independent and uniformly distributed between $(-\pi, \pi]$, and 2) {\em narrow spread of AoAs:} the AoA $\theta_1$ is uniformly distributed between $(-\pi, \pi]$ while $\theta_2$ is uniformly distributed with mean equal to the value of $\theta_1$ and some angular spread (considered to be $10^{\circ}$).}

\begin{theorem} \label{thm:BERExpMO}
{\cc The average BER of Manchester encoding in the MA receiver with uniform spread of the two AoAs is given by}:
\begin{align}
&P_{\rm MA}(e) = \int_{-\pi}^{\pi} \int_{-\pi}^{\pi}\frac{1}{2\pi}\times \frac{1}{2\pi} \times \nonumber\\  
&\!\! \frac{\sigma_n^2}{ G |\alpha|^2 \sigma_h^4 \!\!\left\{\! \mathbb{E} \!\left[ |X\!|^2 \right]  \!\! + \!\! \frac{2\rho_t \rho_b}{1\!-\!\rho_t \rho_b} \!( \!1 \!\!-\!\! \frac{2(\!1- \rho_t^{\frac{N}{2}}\!\! \rho_b^{\frac{N}{2}}\!\!)}{N\!(1\!-\!\rho_t \rho_b)} \!) |\mathbb{E} \!\left[ X \right]\!|^2 \!\! \right\}  \mspace{-3mu} \! +  \mspace{-3mu} 2\sigma_n^2} \mathrm{d} \theta_1 \mathrm{d} \theta_2. \label{eq:bermam2}
\end{align}
\end{theorem}

\begin{IEEEproof}
	See Appendix~\ref{app:BERExpMO}.
\end{IEEEproof} 


\subsubsection*{\textbf{Asymptotic analysis}}
The ratio of the variances of the null and alternate hypotheses of the MA receiver is
\begin{align*}
\!K 
&\!\!= \!1\!\!+\!G |\alpha|^2 \sigma_h^4 \! \left\{\!\! 1 \!\!+\!  \frac{2\rho_t \rho_b}{1\!\!-\!\!\rho_t \rho_b} ( \!1 \!\!-\!\! \frac{2(1\!\!-\!\rho_t^{N/2}\!\! \rho_b^{N/2})}{N(1\!-\!\rho_t \rho_b)}\! ) \frac{ |\mathbb{E}\! \left[ X \right]\!|^2}{\mathbb{E} \!\left[ |X|^2 \right]} \! \!\right\} \!  {\rm SNR}.
\end{align*}
The asymptotic conditional BER of the Manchester encoding in the MA receiver is given by:
\begin{align*}
&P_{\rm MA}^{\rm asym}(e|\phi_1, \phi_2) = \lim_{{\rm SNR} \to \infty} (1+K)^{-1}\nonumber\\
& = \frac{{\rm SNR}^{-1}}{ G |\alpha|^2 \sigma_h^4 \!\!\left\{\! 1\! + \! \frac{2\rho_t \rho_b}{1\!-\!\rho_t \rho_b} ( 1 \!-\! \frac{2(1-\!\rho_t^{\frac{N}{2}}\!\! \rho_b^{\frac{N}{2}}\!\!)}{N(1\!-\!\rho_t \rho_b)} ) \frac{ |\mathbb{E} \left[ X \right]|^2}{\mathbb{E} \left[ |X|^2 \right]} \!\! \right\} \! + \! 2 {\rm SNR}^{\!-1}} \!=\! 0. \label{eq:asymBERM1MA}
\end{align*}

%

\section{Numerical Results and Discussion} \label{sec:NumResults}

\begin{figure}
    \centering
    \begin{subfigure}[b]{0.45\textwidth}
        \centering
        \includegraphics [width=\linewidth]{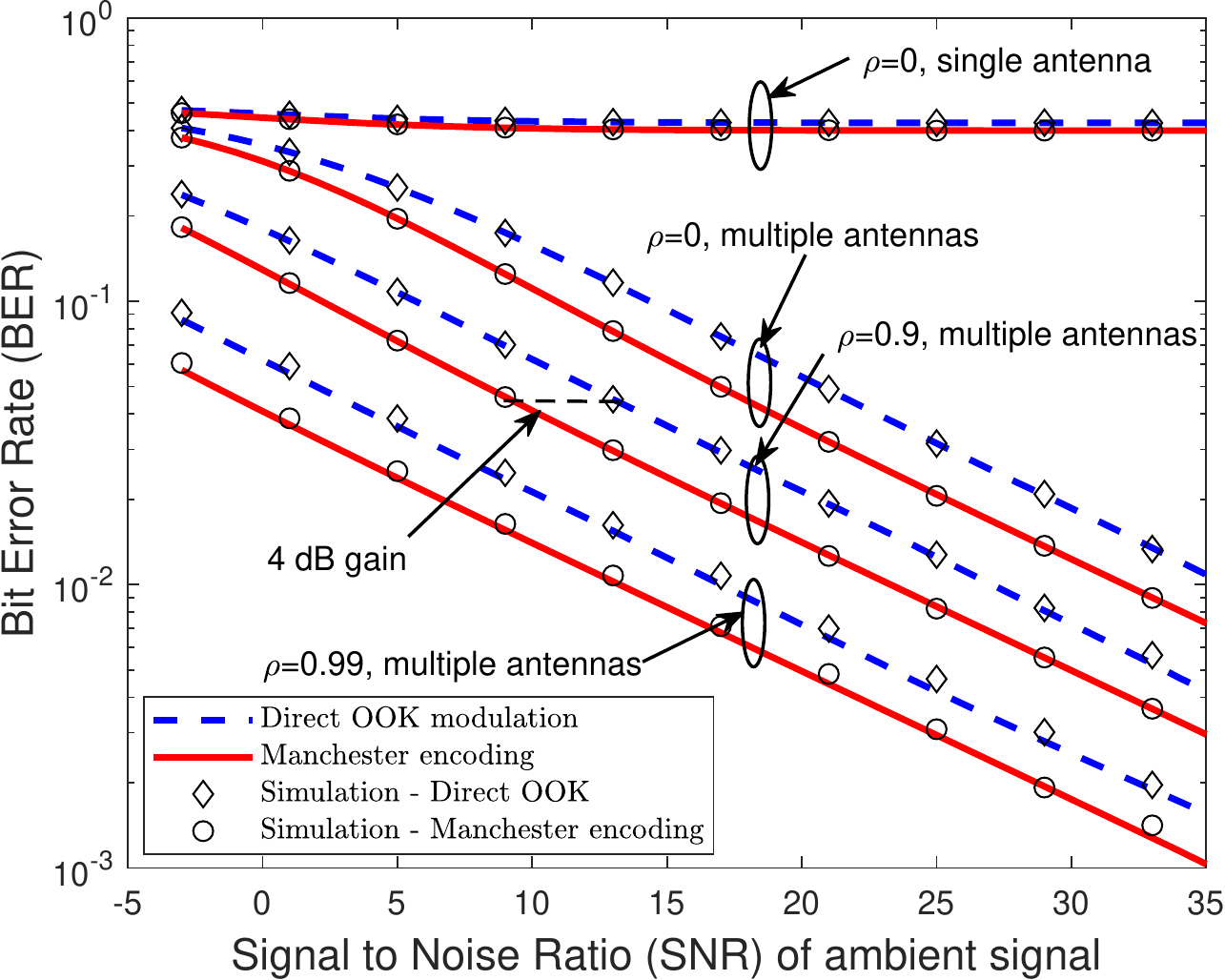}
         \caption{}\label{fig:BER_FS_manchester_coding_comp_SNR}
    \end{subfigure}
    ~ 
    \begin{subfigure}[b]{0.45\textwidth}
        \centering
        \includegraphics [width=\linewidth]{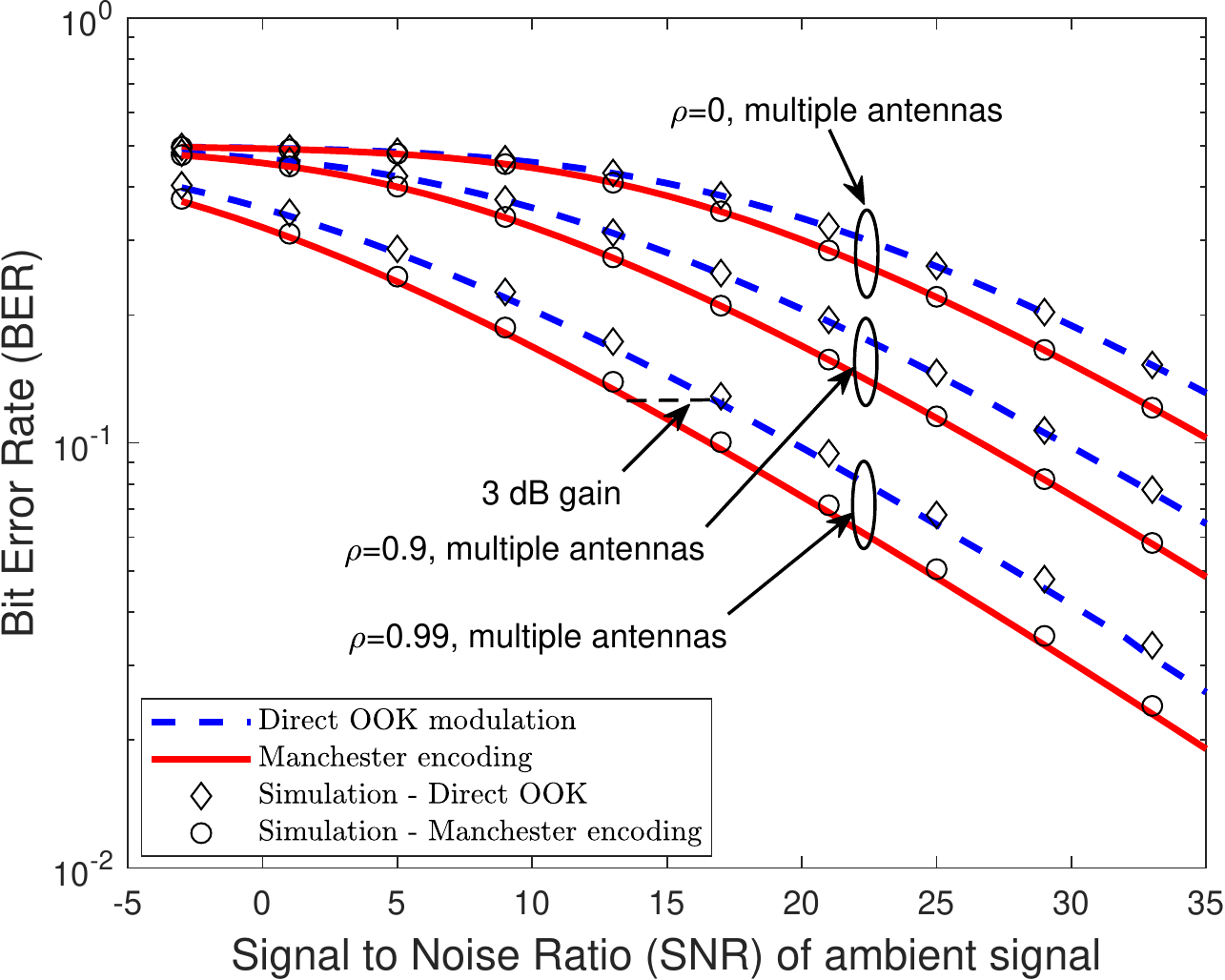}
	\caption{}\label{fig:BER_FS_manchester_coding_comp_10d_SNR}
    \end{subfigure}
    \caption{\cc BER vs SNR comparison of Manchester encoding and the direct OOK modulation for varying correlation factor $\rho$, $M_r=4$ and $N=2000$: (a) uniform spread of AoAs, (b) narrow spread of AoAs.}
\end{figure}

\begin{figure}
    \centering
    \begin{subfigure}[b]{0.45\textwidth}
        \centering
        \includegraphics [width=\linewidth]{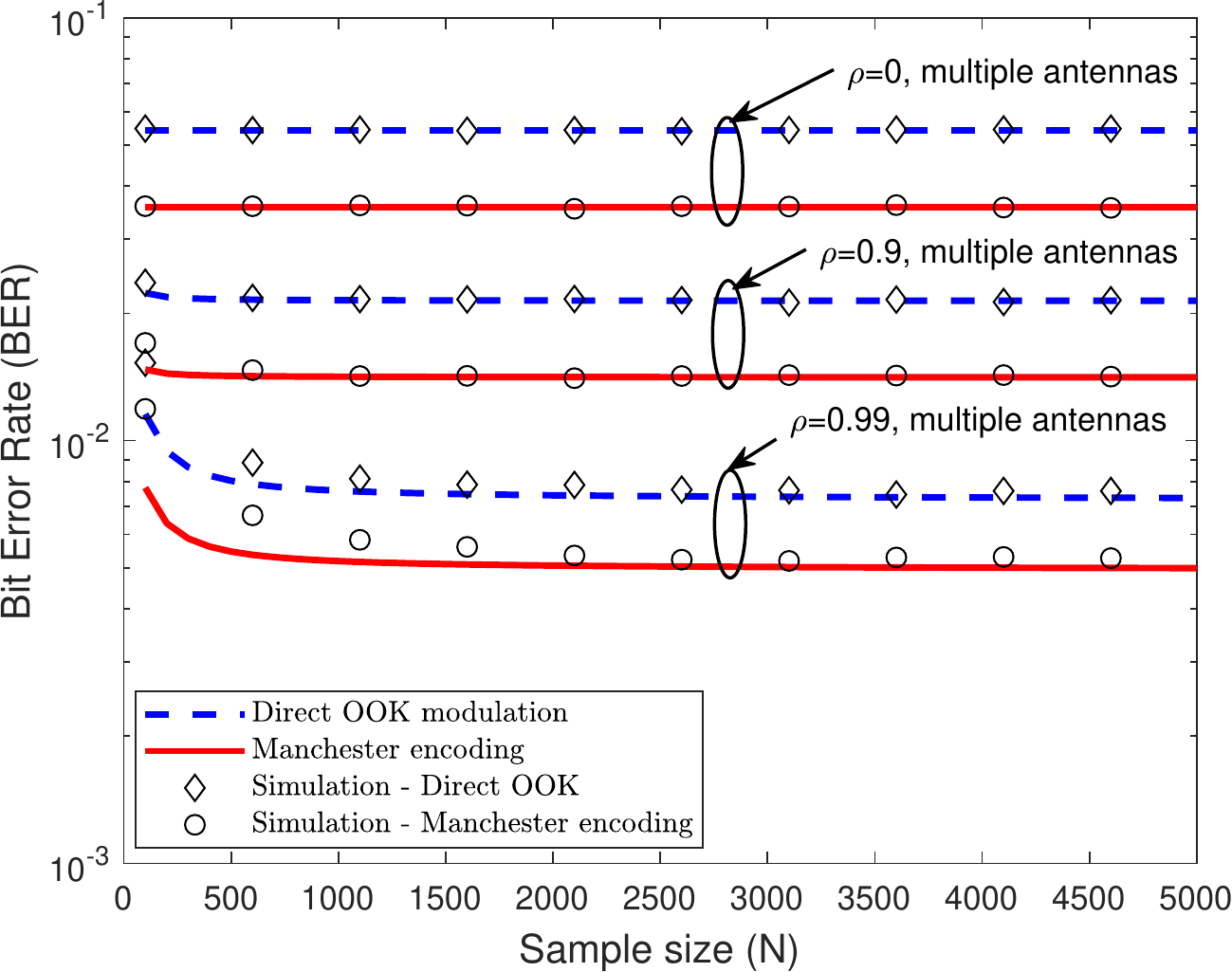}
         \caption{}\label{fig:BER_FS_manchester_coding_comp_N}
    \end{subfigure}
    ~ 
    \begin{subfigure}[b]{0.45\textwidth}
        \centering
        \includegraphics [width=\linewidth]{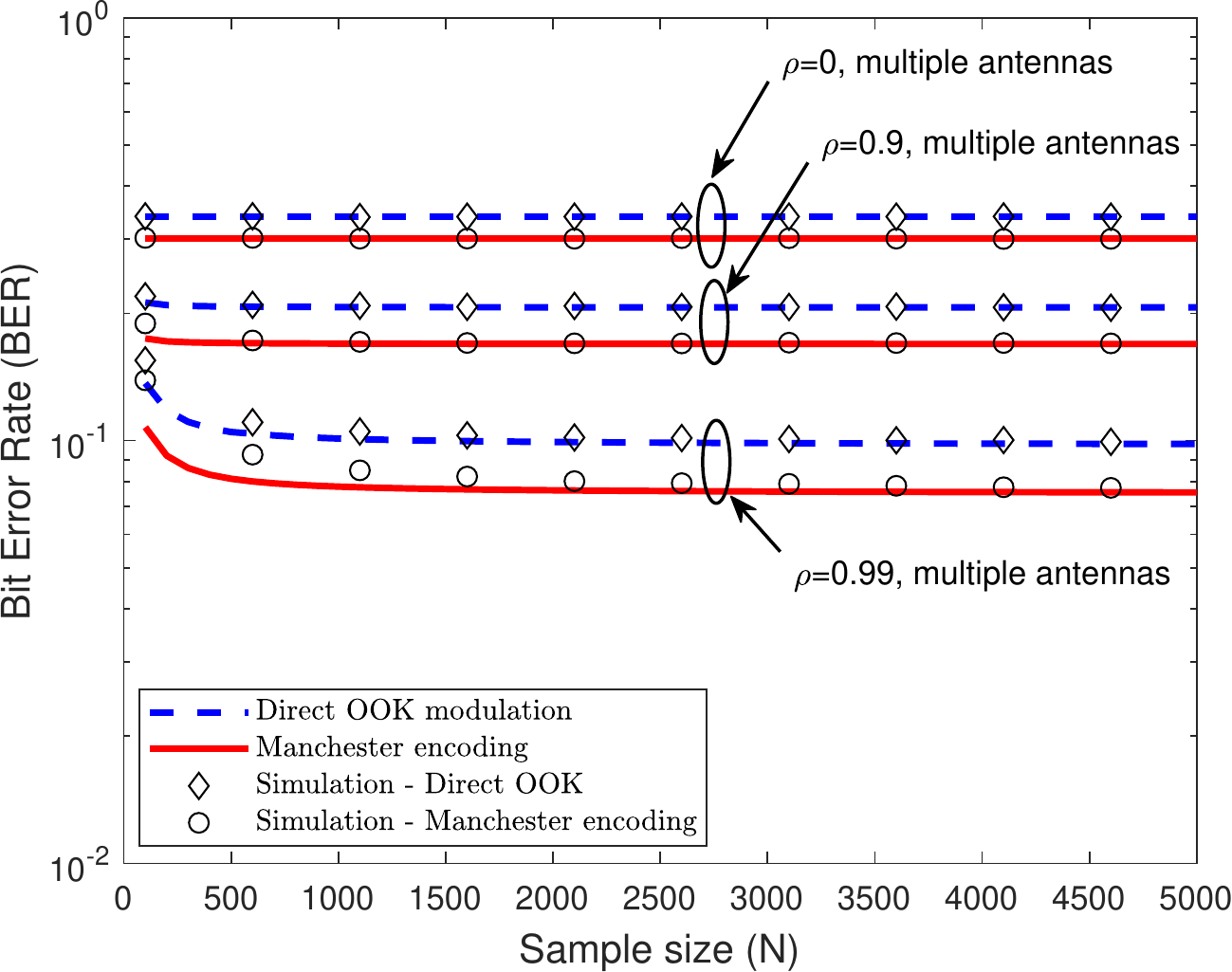}
	\caption{}\label{fig:BER_FS_manchester_coding_comp_10d_N}
    \end{subfigure}
    \caption{\cc BER vs N comparison of Manchester encoding and the direct OOK modulation for varying correlation factor $\rho$, $M_r=4$, and $ {\rm SNR}=20 \,{\rm dB}$: (a) uniform spread of AoAs, (b) narrow spread of AoAs.}
\end{figure}

We now compare the BER performance of Manchester encoding with the direct OOK modulation. In addition, the analytical results are compared with Monte-Carlo simulation to verify the accuracy of our analysis. The value of $\alpha$ is configured to result in a signal attenuation of 1.1 dB, while the variance of the channel gains $\sigma_{h}^2$ is set to $1$. The values of all the correlation factors $\rho_r, \rho_b$ and $\rho_t$ are assumed to be the same, and are represented using another variable $\rho$. {\cc First, we discuss the average BER results for the uniformly spread AoAs before comparing with the performance under narrowly spread AoAs. The BER result of the uncoded and coded schemes of the SA receiver for the independent fading scenario ($\rho=0$) is shown in Fig. \ref{fig:BER_FS_manchester_coding_comp_SNR}, and it can be verified from the plot that both of the  schemes suffer from an error floor which can be attributed to the DL interference. In the same Fig. \ref{fig:BER_FS_manchester_coding_comp_SNR}, the BER results of the MA receiver for the two schemes with varying values of the correlation factor $\rho$ are also plotted. The effect of the DL cancellation on the performance can be inferred from the improved BER of the two schemes. It can  also be verified from the figure that Manchester encoding results in an SNR gain of around $4$ dB over the uncoded direct OOK modulation for the uniformly spread AoAs. In comparison, the SNR gain of the Manchester encoding over the uncoded OOK modulation obtained for the narrowly spread AoAs is around $3$ dB, as shown in Fig. \ref{fig:BER_FS_manchester_coding_comp_10d_SNR}. As expected, the exact SNR gain with the Manchester encoding is dependent on the joint distribution of the two AoAs but remains constant for different values of $\rho$. In addition, the BER curves of the two schemes with increasing sample-size $N$ for the uniformly spread and narrow spread AoAs are respectively plotted in Figs. \ref{fig:BER_FS_manchester_coding_comp_N} and \ref{fig:BER_FS_manchester_coding_comp_10d_N}, which are flat beyond a threshold value of $N$. The mismatch between the theoretical and simulation results for small $N$ is due to the requirement of minimum number of samples for the averaging operation to work properly, and larger number of samples  are required with increasing value of the correlation factor $\rho$.}

\section{Conclusion} \label{sec:conclusion}

In this work, we have analyzed the impact of Manchester encoding on non-coherent transmission with respect to an ambient backscatter system under time-selective fading, where we have analytically and numerically shown the advantages of the scheme over the conventional direct OOK modulation used in the literature. The optimal decision rule for  Manchester encoding is only dependent on the relative magnitude of the test statistic for the two symbols of the codeword, and hence the optimal detection threshold 
turns out to be independent of all the system parameters. {\cc In addition, the proposed encoding scheme also achieves an SNR gain over the direct OOK modulation with the MA receiver, the exact value of which will vary based on the joint distribution of the two AoAs. In our analysis, the SNR gain evaluated for the uniformly and narrowly spread AoAs came  to be around $4$ dB and $3$ dB, respectively, which is a substantial improvement in the performance of the ambient backscatter system.}

\appendix

\subsection{Proof of Lemma~\ref{lem:CondPDF}} \label{app:CondPDF}

The test statistics $Z_0$ and $Z_1$ are correlated due to the common signal component from the DL present in the two codeword symbols. And, since they are jointly Gaussian, deriving covariance for the two symbols in addition to their individual variances is sufficient. The joint distribution of the bivariate Gaussian random variables is given by:
\begin{align}
f_{Z_0, Z_1}\left( z_0, z_1 \right) = \frac{1}{\pi^2 |C_z|^{1/2}} e^{-\frac{1}{2} (z-m)^H C_z^{-1} (z-m)},
\end{align}
where the mean in this case is $m = E[Z_0 Z_1] = \bar{0}$ since the channel is Rayleigh faded, and from this the covariance matrix also simplifies as follows:
\begin{align*}
C_z = \begin{bmatrix}
   {\rm Var}[Z_0]& \bar{{\rm Cov}}[Z_0, Z_1]\\
   {\rm Cov}[Z_0, Z_1] & {\rm Var}[Z_1]
  \end{bmatrix}.
\end{align*}
 The covariance is non-zero as a result of the DL present in the two consecutive symbols, and it can be easily verified that ${\rm Cov}[Z_0, Z_1] =\bar{{\rm Cov}}[Z_0, Z_1]$ from the symmetry of the problem (and therefore real). Due to this symmetry, it is enough to evaluate ${\rm Cov}[Z_0, Z_1]$ under null hypothesis $\mathcal{H}_0$. The conditional covariance of any two samples $y[i]$ and $y[j]$, for $j>i$, is given by:
\begin{align*}
&{\rm Cov}\left[y[i], y[j] \right] \stackrel{(a)}{=}  \mathbb{E} \left[y[i] y^*[j] \right] \\
&\!\!=\! \mathbb{E} \!\Bigg[\!(\rho_r^{i\!+\!j\!-\!2} |h_r[1]|^2  \!\!+ \!\!\sqrt{1\!\!-\!\!\rho_r^2} \!\!\sum\limits_{k_2=1}^{j-1} \!\! \rho_r^{i\!+j-\!k_2-\!2}h_r[1]g_r^*[k_2] ) x[i] x^*[j]\\
& + \rho_r^{i-1} h_r[1] x[i] w^*[j] \! +\! ( \sqrt{1-\rho_r^2} \sum\limits_{k_1=1}^{i-1}\rho_r^{i+j-k_1-2}h_r^*[1]g_r[k_1]  \\
&+ (1-\rho_r^2) \sum\limits_{k_1=1}^{i-1} \sum\limits_{k_2=1}^{j-1}\rho_r^{i+j-k_1-k_2-2}g_r[k_1]g_r^*[k_2] ) x[i] x^*[j] \nonumber\\
& \!\!+\!\! \sqrt{1-\rho_r^2} \sum\limits_{k_1=1}^{i-1}\rho_r^{i-k_1-1}w^*[j]g_r[k_1] x[i]\! +\! \rho_r^{j-1} h_r[1] x^*[j] w[i]  \nonumber\\
& + \sqrt{1-\rho_r^2} \sum\limits_{k_2=1}^{j-1}\rho_r^{j-k_2-1}w[i]g_r^*[k_2] x^*[j] + w[i] w^*[j]\Bigg]\\
&\!\!= \! \sigma_h^2 \!(\rho_r^{i+j-2} \!\! +\!\! (1\!-\!\rho_r^2)\!\! \sum\limits_{k=1}^{i-1} \! \rho_r^{i+j-2k-2}) x[i] x^{\!*}[j]\!\! =\!\! \sigma_h^2 \rho_r^{j-i} \!x[i] x^{\!*}[j],
\end{align*}
where (a) follows from the fact that the conditional expectation of the samples is zero.  The covariance ${\rm Cov}^{\rm SA}$ can be evaluated as:
\begin{align*}
&{\rm Cov}^{\rm SA} = {\rm Cov}[Z_0, Z_1] 
= \frac{4}{N^2} ( \sum\limits_{n_1, n_2} {\rm Cov}\left[y[n_1], y[n_2] \right]) \nonumber\\
&\!\!=\!\! \frac{4\sigma_h^2}{N^2}\!\! \sum\limits_{n_1=1}^{N/2} \sum\limits_{n_2=N/2+1}^{N}\!\!\!\!\!\!\! \rho_r^{n_2-n_1}  x[n_1] x^*[n_2] \!=\!\! \frac{4\rho_r \!(\!1\!\!-\!\!\rho_r^{N/2}\!)^2}{N^2 \!\left( \!1\!\!-\!\!\rho_r\!\right)^2}  \!|\mathbb{E}\! \left[ X \right]\!|^2\!. 
\end{align*}
For the null hypothesis $\mathcal{H}_0$, the variances ${\rm Var}[Z_0]$ and ${\rm Var}[Z_1]$ are given by ${\rm Var}^{\rm SA}_0$  and ${\rm Var}^{\rm SA}_1$, respectively, whose 
derivations follow a procedure similar to the one used for ${\rm Cov}^{\rm SA}$. These derivations are, therefore, skipped here due to space constraints, and interested readers can refer to \cite[Lemmas 2 and 3]{devineni2019non} for the details. The variances of $\mathcal{H}_0$ are simply exchanged to get the respective variances of $Z_0$ and $Z_1$ under the alternate hypothesis $\mathcal{H}_1$.

\subsection{Proof of Theorem~\ref{thm:BERExpSO}} \label{app:BERExpSO}

Comparing the joint conditional PDFs derived in Lemma~\ref{lem:CondPDF}, the  optimal decision rule can be obtained as:
\begin{align}
	&  \ln \left[ f_{Z_0, Z_1 | \mathcal{H}_0 }\left( z_0, z_1 \right) \right] \gtrless_{1}^{0} \ln \left[ f_{Z_0, Z_1 | \mathcal{H}_1 }\left( z_0, z_1 \right) \right] \nonumber\\
	&  \implies -\left|z_0\right|^2 {\rm Var}^{\rm SA}_1  - \left|z_1\right|^2 {\rm Var}^{\rm SA}_0 + (z_0 z_1^*+ z_0^* z_1) {\rm Cov}^{\rm SA} \nonumber\\
	&  \gtrless_{1}^{0} - \left|z_0\right|^2 {\rm Var}^{\rm SA}_0  - \left|z_1\right|^2 {\rm Var}^{\rm SA}_1 + (z_0 z_1^*+ z_0^* z_1) {\rm Cov}^{\rm SA} \nonumber\\
	& \quad \quad \quad \quad \,\,\, \implies |z_0|^2 \gtrless_{0}^{1} |z_1|^2 \label{eq: optdecrule}.
\end{align} 
Since the two hypotheses are symmetric, it is sufficient to evaluate the average BER for the null hypothesis $\mathcal{H}_0$, which is evaluated using the joint PDF as follows:
\begin{align*}
&P_{\rm SA}(e) \!\! = \!\! Pr\left\{|Z_0|^2 \!>\! |Z_1|^2 \!\mid \! \mathcal{H}_0\right\} \!\! = \!\! \int\limits_{0}^{\infty} \!\! \int\limits_{v}^{\infty} \!\! f_{|Z_0|^2, |Z_1|^2}\left( u,v \right) \,\mathrm{d}u \,\mathrm{d}v\\
&\stackrel{(a)}{=}  \int_{0}^{\infty} \int_{v}^{\infty} \frac{\exp\left\{-\left( \frac{u}{( 1 - \rho^2){\rm Var}[Z_0]}  + \frac{v}{( 1 - \rho^2){\rm Var}[Z_1]} \right)\right\}}{\pi ( 1 - \rho^2){\rm Var}[Z_0]{\rm Var}[Z_1]} \nonumber\\
& \quad \quad \quad \quad \times I_0(\frac{\rho \sqrt{u v}}{( 1 - \rho^2)\sqrt{{\rm Var}[Z_0]{\rm Var}[Z_1]}}) \,\mathrm{d}u \,\mathrm{d}v,
\end{align*}
where $(a)$ results from the fact that the joint distribution of magnitude squares of the bi-variate Gaussian random variables is characterized as a bi-variate Rayleigh \cite{middleton1960introduction}.

From the expressions given for ${\rm Var}^{\rm SA}_0, {\rm Var}^{\rm SA}_1$ and ${\rm Cov}^{\rm SA}$ in Lemma \ref{lem:CondPDF}, one can conclude that the covariance ${\rm Cov}^{\rm SA}$ decays faster compared to the variances ${\rm Var}^{\rm SA}_0$  and ${\rm Var}^{\rm SA}_1$. Hence, for a sufficiently large value of the sample-size $N$, the random variables $|Z_0|^2$ and $|Z_1|^2$ can be approximated as independent. Consequently, the joint distribution of $|Z_0|^2$ and $|Z_1|^2$ simplifies to the product of their marginal distributions. Note that the marginal PDFs of $|Z_0|^2$ and $|Z_1|^2$ are exponential. Due to the symmetry present in the two hypotheses of the problem, we only need to derive the error probability for $\mathcal{H}_0$. The derivation for the theoretical average BER of the SA receiver is given as follows:
\begin{align*}
&P_{\rm SA}(e) \!= \!Pr\!\left\{\!|Z_0|^2 \!> \! |Z_1|^2 \!\mid \!\mathcal{H}_0 \!\right\} \!\!=\!\! Pr\!\left\{\!|Z_0|^2 \! > \! t \! \mid \! |Z_1|^2 \!=\! t, \mathcal{H}_0 \!\right\}\\
&= \int_{0}^{\infty}\left[1- F_{\rm Exp}\left(t,{\rm Var}^{\rm SA}_0\right) \right] f_{\rm Exp}\left(t, {\rm Var}^{\rm SA}_1 \right) \,\mathrm{d}t\\
&= \int_{0}^{\infty} e^{{-\frac{t}{{\rm Var}^{\rm SA}_0}}} \frac{e^{{-\frac{t}{{\rm Var}^{\rm SA}_1}}}}{{\rm Var}^{\rm SA}_1}\, \mathrm{d}t  =   \int_{0}^{\infty} \frac{ e^{-t\left(\frac{1}{{\rm Var}^{\rm SA}_0} + \frac{1}{{\rm Var}^{\rm SA}_1}\right)}}{{\rm Var}^{\rm SA}_1} \, \mathrm{d}t \\
& =  \left(1+ \frac{{\rm Var}^{\rm SA}_1}{{\rm Var}^{\rm SA}_0}\right)^{\!\!\!-1}\!\!\!\!\!,
\end{align*}
where $F_{\rm Exp}(x,\lambda)$ and $f_{\rm Exp}(x,\lambda)$ are the cumulative distribution function and the PDF of an exponential RV with mean $\lambda$, respectively. 
 
\subsection{Proof of Theorem~\ref{thm:BERExpMO}} \label{app:BERExpMO}
Since the DL is canceled in the MA receiver, no correlation exists between the two variables $Z_0$ and $Z_1$ of the codeword. Hence, the conditional joint PDFs of $Z_0$ and $Z_1$ are given by:
\begin{align*}
\mathcal{H}_0 \!\!\begin{cases}
Z_0 \sim \mathcal{CN} \left(0,{\rm Var}^{\rm MA}_0\right)\\
Z_1 \sim \mathcal{CN}\left(0, {\rm Var}^{\rm MA}_1\right),
\end{cases} &  \mathcal{H}_1 \!\! \begin{cases}
Z_0 \sim \mathcal{CN} \left(0, {\rm Var}^{\rm MA}_1\right)\\
Z_1 \sim \mathcal{CN}\left(0,{\rm Var}^{\rm MA}_0\right),
\end{cases}
\end{align*}
where ${\rm Var}^{\rm MA}_1 \!\!= \!\!\frac{ G |\alpha|^2 \sigma_h^4 \left\{ \! \mathbb{E} \left[ |X|^2 \right] + \! \frac{2\rho_t \rho_b}{1-\rho_t \rho_b} \left( \!\!1 -\! \frac{1-\rho_t^{N}\!\rho_b^{N}}{N(1-\rho_t \rho_b)} \!\!\right) |\mathbb{E} \left[ X \right]|^2 \! \right\}+ \sigma_n^2}{N}$ and ${\rm Var}^{\rm MA}_0 \!\!=\!\! \frac{ \sigma_n^2}{N}$ are the variances of the MA receiver as derived in \cite{devineni2019non} for the direct OOK modulation. The optimal decision rule once again turns out to be~\eqref{eq: optdecrule}, from which the conditional BER can be derived as:
\begin{align}
&P(e|\phi_1, \phi_2) =  \left(\!1\!+\! \frac{{\rm Var}^{\rm MA}_1}{{\rm Var}^{\rm MA}_0}\!\right)^{\!\!-1} \nonumber\\
&= \frac{\sigma_n^2}{ G |\alpha|^2 \sigma_h^4 \!\left\{\! \mathbb{E} \!\left[ |X|^2 \right] \! + \! \frac{2\rho_t \rho_b}{1\!-\!\rho_t \rho_b} ( \!1 \!-\! \frac{2(\!1-\rho_t^{\frac{N}{2}}\!\! \rho_b^{\frac{N}{2}}\!)}{N\!(1-\rho_t \rho_b)} ) |\mathbb{E} \!\left[ X \right]\!|^2 \! \right\}   \! +\! 2 \sigma_n^2} . \nonumber
\end{align}
Since the antenna gain depends on the phase offsets of the two links (and thereby their AoAs), the average BER is derived by marginalizing over the range $(-\pi, \pi]$ of the AoAs $\theta_1$ and $\theta_2$.

{ 
\bibliographystyle{IEEEtran}
	\bibliography{ambnoncoherent}

\begin{thebibliography}{10}
\providecommand{\url}[1]{#1}
\csname url@samestyle\endcsname
\providecommand{\newblock}{\relax}
\providecommand{\bibinfo}[2]{#2}
\providecommand{\BIBentrySTDinterwordspacing}{\spaceskip=0pt\relax}
\providecommand{\BIBentryALTinterwordstretchfactor}{4}
\providecommand{\BIBentryALTinterwordspacing}{\spaceskip=\fontdimen2\font plus
\BIBentryALTinterwordstretchfactor\fontdimen3\font minus
  \fontdimen4\font\relax}
\providecommand{\BIBforeignlanguage}[2]{{%
\expandafter\ifx\csname l@#1\endcsname\relax
\typeout{** WARNING: IEEEtran.bst: No hyphenation pattern has been}%
\typeout{** loaded for the language `#1'. Using the pattern for}%
\typeout{** the default language instead.}%
\else
\language=\csname l@#1\endcsname
\fi
#2}}
\providecommand{\BIBdecl}{\relax}
\BIBdecl

\bibitem{devineni2019ambient}
J.~K. Devineni and H.~S. Dhillon, ``Ambient backscatter systems: Exact average
  bit error rate under fading channels,'' \emph{IEEE Trans. Green Commun. and
  Networking}, vol.~3, no.~1, pp. 11--25, Mar. 2019.

\bibitem{gao17}
J.~Qian, F.~Gao, G.~Wang, S.~Jin, and H.~Zhu, ``Noncoherent detections for
  ambient backscatter system,'' \emph{IEEE Trans. Wireless Commun.}, vol.~16,
  no.~3, pp. 1412--1422, Mar. 2017.

\bibitem{gao16}
------, ``Semi-coherent detection and performance analysis for ambient
  backscatter system,'' \emph{IEEE Trans. Commun.}, vol.~65, no.~12, pp.
  5266--5279, Dec. 2017.

\bibitem{el2019noncoherent}
M.~A. El~Mossallamy, M.~Pan, R.~J{\"a}ntti, K.~G. Seddik, G.~Y. Li, and Z.~Han,
  ``Noncoherent backscatter communications over ambient {OFDM} signals,''
  \emph{IEEE Trans. Commun.}, vol.~67, no.~5, pp. 3597 -- 3611, May 2019.

\bibitem{elmossallamy2019noncoherent}
M.~ElMossallamy, Z.~Han, M.~Pan, R.~Jantti, K.~Seddik, and G.~Y. Li,
  ``Noncoherent frequency shift keying for ambient backscatter over {OFDM}
  signals,'' \emph{Proc., IEEE ICC}, May 2019.

\bibitem{zhao2018blind}
W.~Zhao, G.~Wang, S.~Atapattu, and B.~Ai, ``Blind channel estimation in ambient
  backscatter communication systems with multiple-antenna reader,''
  \emph{Proc., IEEE/CIC ICCC}, pp. 320--324, Aug. 2018.

\bibitem{guo2018noncoherent}
H.~Guo, Q.~Zhang, D.~Li, and Y.-C. Liang, ``Noncoherent multiantenna receivers
  for cognitive backscatter system with multiple {RF} sources,'' \emph{arXiv
  preprint, arXiv:1808.04316}, 2018.

\bibitem{tao2018symbol}
Q.~Tao, C.~Zhong, H.~Lin, and Z.~Zhang, ``Symbol detection of ambient
  backscatter systems with manchester coding,'' \emph{IEEE Trans. Wireless
  Commun.}, vol.~17, no.~6, pp. 4028--4038, 2018.

\bibitem{zhao2019channel}
W.~Zhao, G.~Wang, S.~Atapattu, R.~He, and Y.-C. Liang, ``Channel estimation for
  ambient backscatter communication systems with massive-antenna reader,''
  \emph{IEEE Trans. on Veh. Tech.}, vol.~68, no.~8, pp. 8254--8258, Aug. 2019.

\bibitem{devineni2019non}
J.~K. Devineni and H.~S. Dhillon, ``Non-coherent detection and bit error rate
  for an ambient backscatter link in time-selective fading,'' \emph{IEEE Trans.
  Commun.}, vol.~69, no.~1, pp. 602--618, Jan. 2021.

\bibitem{stuber2017principles}
G.~St{\"u}ber, \emph{Principles of Mobile Communication}.\hskip 1em plus 0.5em
  minus 0.4em\relax Springer International Publishing, 2017.

\bibitem{liu2002space}
Z.~Liu, X.~Ma, and G.~B. Giannakis, ``Space-time coding and {K}alman filtering
  for time-selective fading channels,'' \emph{IEEE Trans. Commun.}, vol.~50,
  no.~2, pp. 183--186, 2002.

\bibitem{devineni19noncoherent}
J.~K. Devineni and H.~S. Dhillon, ``Non-coherent signal detection and bit error
  rate for an ambient backscatter link under fast fading,'' \emph{Proc., IEEE
  Globecom}, Dec. 2019.

\bibitem{middleton1960introduction}
D.~Middleton, \emph{An introduction to statistical communication theory}.\hskip
  1em plus 0.5em minus 0.4em\relax McGraw-Hill New York, 1960, vol. 960.

\end{thebibliography}
}

\end{document}